\documentclass[conference]{IEEEtran}
\usepackage{amsmath}
\usepackage[pdftex]{graphicx}
\usepackage{graphicx}
\usepackage{pstricks}
\usepackage{amsfonts}
\usepackage{tabularx}

\newtheorem{Remark}{Remark}

\usepackage{amssymb}
\usepackage{multirow}
\usepackage{epsfig}
\usepackage{epstopdf}

\usepackage[noadjust]{cite} 

\begin{document}
\title{Secrecy Outage Probability Analysis for Downlink Untrusted NOMA Under Practical SIC Error}
\author{\IEEEauthorblockN{Sapna Thapar$^{1}$, Deepak Mishra$^{2}$, Derrick Wing Kwan Ng$^{2}$, and Ravikant Saini$^{1}$}
\IEEEauthorblockA{$^{1}$Department of Electrical Engineering, Indian Institute of Technology Jammu, India\\
$^{2}$School of Electrical Engineering and Telecommunications, University of New South Wales, Australia \\
Emails: thaparsapna25@gmail.com, d.mishra@unsw.edu.au, w.k.ng@unsw.edu.au,  ravikant.saini@iitjammu.ac.in
}
}
\maketitle

\begin{abstract}
Non-orthogonal multiple access (NOMA) serves multiple users simultaneously via the same resource block by exploiting superposition coding at the transmitter and successive interference cancellation (SIC) at the receivers. Under practical considerations, perfect SIC may not be achieved. Thus, residual interference (RI) occurs inevitably due to imperfect SIC. 
In this work, we first propose a novel model for characterizing RI to provide a more realistic secrecy performance analysis of a downlink NOMA system under imperfect SIC at receivers. In the presence of untrusted users, NOMA has an inherent security flaw. Therefore, for this untrusted users' scenario, we derive new analytical expressions of secrecy outage probability (SOP) for each user in a two-user untrusted NOMA system by using the proposed RI model. To further shed light on the obtained results and obtain a deeper understanding, a high signal-to-noise ratio approximation of the SOPs are also obtained. Lastly, numerical investigations are provided to validate the accuracy of the desired analytical results and present valuable insights into the impact of various system parameters on the secrecy rate performance of the secure NOMA communication system.
\end{abstract}

\section{Introduction and Background}\label{section1}
Non-orthogonal multiple access (NOMA) has been regarded as a disruptive technology to satisfy the requirement of high spectral efficiency for fifth-generation and beyond wireless networks \cite{7263349}. 
From the security perspective, sharing the same time-frequency resource among users in NOMA imposes secrecy challenges if the multiplexed users are untrusted. Recently, physical layer security (PLS) has emerged as a new research frontier among researchers for securing communication confidentiality via exploiting the random nature of wireless transmission media \cite{wyner1975wire}. Therefore, achieving secure NOMA communication in the presence of untrusted users by utilizing the potential of PLS is a promising research area. 

\subsection{State-of-the-Art}\label{literature_survey}
A system with untrusted users is a hostile scenario where users do not trust each other and focus on safeguarding their own data from all other users by treating them as eavesdroppers. 
In this regard,  for a two-user NOMA system, the secrecy outage probability (SOP) of a trusted strong user against an untrusted weak user was analyzed in \cite{basepaper}. In \cite{8945391}, two optimal relay selection schemes were presented to achieve reliable and secure communication for a strong user against an untrusted weak user. 
A PLS technique was presented in \cite{9004475} to secure the data of a weak user from an untrusted strong user by using a directional demodulation approach. 
A linear precoding technique was proposed in \cite{9169675} to secure each  user's data from its counterpart. In  \cite{9188014}, an optimal decoding order was proposed with respect to providing a positive secrecy rate for both unttusted users against each other in a two-user NOMA system. An $N$-users untrusted NOMA system was investigated from the perspective of decoding orders in \cite{9324786}. 

Despite the fruitful results in the literature in handling untrusted users in NOMA systems, most of the existing works, e.g. \cite{basepaper}-\cite{9169675}, over-optimistically assumed that the receivers are capable of performing perfect successive interference cancellation (SIC). 
Even though better spectral efficiency can be attained if perfect SIC can be performed,  it might not be realizable in practice due to several implementation issues such as decoding error and complexity scaling \cite{7881111}. Consequently, imperfect SIC, where the residual interference (RI) arising from incorrectly decoded users' messages inevitably remains and accumulates, should be considered in the system model. Recently, numerous researchers have analyzed the performance of NOMA with imperfect SIC \cite{7881111}, \cite{8125754}, \cite{constant_sic1}, \cite{constant_sic2}.  However, in untrusted secure NOMA, a few research works, e.g., \cite{9188014}, \cite{9324786}, have considered the impact of RI on receivers but with different drawbacks, as will be discussed in the following.

\subsection{Motivation and Contributions}\label{research_gap}
Some of the existing works, e.g., \cite{9188014}, \cite{constant_sic1}, \cite{constant_sic2}, considering  NOMA with imperfect SIC, have assumed a certain constant value of RI at receivers. We refer to this case as the \emph{constant model}. 
This strong unrealistic assumption not only overly simplifies the model but also leads to prediction errors. 
Besides, most of the existing works have modeled the RI as a fixed fraction of the power of the signal, which is aimed to be cancelled \cite{9324786}-\cite{8125754}. We refer to this case as the \emph{fixed model}. However, in practice, the actual impacts of imperfect SIC cannot be fully revealed if a fixed model is adopted. In fact, via experimental and theoretical investigations for packet transmission networks, it has been demonstrated in \cite{7842164} that the average realistic RI is not a constant fraction of the power of the packet being cancelled. 

Note that in practical NOMA systems, the performance of decoders significantly depend on the signal-to-interference-noise ratio (SINR) of the received signal and the target SINR requirement of each user \cite{8171091}. Many existing works have considered that each user may have a predefined threshold requirement, and successful decoding is possible if the achievable SINR is higher than the threshold  \cite{9151196}, \cite{8309422}. Although these works have considered the SINR criterion to determine whether successful decoding is possible or not, they have not determined the RI occurred due to imperfect SIC in terms of achieved SINR at decoded
users. Thus, the motivation of this work is to investigate the realistic impact of SIC imperfections. 
The key contributions are summarized as follows: 
\begin{itemize}
\item  
We present a thorough analysis of imperfect SIC in practical receivers, and propose a novel RI model that characterizes the RI based on the SINR achieved in the previous stages for decoding the interfering signal's data.
\item Considering the proposed RI model in a two-user untrusted NOMA system, the analytical expressions of SOP are derived to analyze secrecy performance at each user against its counterpart. 
\item The SOPs have been further investigated under high SNR scenarios, and asymptotic closed-form approximations of SOPs are obtained. 
\item Numerical results are provided to validate the derived analytical expressions presenting valuable insights on the impact of various key parameters on system performance.
\end{itemize} 



\section{NOMA Transmission among Untrusted Users}\label{section2}

\subsection{System Model}\label{system_model}
We consider a downlink power-domain NOMA system with one base station (BS) and two untrusted users. 
We denote the $n$-th user by $U_{n}$, where $n\in \mathcal{N}=\{1, 2\}$. Each node in the network has a single antenna \cite{basepaper}, \cite{9188014}. The channel gain coefficient between the BS and $U_{n}$, denoted as $h_{n}$, is considered to be Rayleigh faded \cite{basepaper}, \cite{9188014}, \cite{7117391}. The channel power gain $|h_{n}|^{2}$ follows exponential distribution with mean parameter $\lambda_{n}=L_{c}d_{n}^{-e}$, where $d_{n}$, $L_{c}$, and $e$, respectively, imply distance between BS and $U_{n}$, path loss constant, and path loss exponent. Without loss of generality, we assume that $|h_{1}|^{2}>|h_{2}|^{2}$. Thus, depending on the channel power gain conditions, $U_{1}$ and $U_{2}$ are, respectively, regarded as strong and weak users.
We denote the total power transmitted from BS to both users by $P_{t}$. The power allocation (PA) coefficient that denotes the fraction of $P_{t}$ allocated to $U_{1}$ is symbolized by $\alpha$. The remanent fraction $(1-\alpha)$ is alloted to $U_{2}$.  Without loss of generality, the receiver noise is assumed to be additive white Gaussian with mean $0$ and variance $\sigma^{2}$ at both the users. 

\subsection{Possible Decoding Orders and Achievable Secrecy Rates}\label{DO}
Unlike conventional decoding order, each user can perform SIC \cite{7343355} and  decode data of itself and other users in any sequence if the users are untrusted \cite{9324786}. Thus, for a NOMA system with $2$ untrusted users, the total possible decoding orders are $4$ as given in \cite{9188014}, \cite{9324786}. 
We represent the decoding order as a $2\times 2$ matrix $\mathbf{D}_{o}$, where $o \in \{1,2,3,4\}$ is an index pointing the $o$-th decoding order. We denote $m$-th column of the matrix by a column vector $\mathbf{d}_{m}$ of size $2\times1$, demonstrating the SIC sequence utilized by $U_{m}$, where $m \in \mathcal{N}$.  Particularly, $[\mathbf{d}_{m}]_{k}=n$ represents that at the $k$-th stage, $U_{m}$ decodes data of $U_{n}$, where $n, k \in \mathcal{N}$ and $[\mathbf{d}_{m}]_{1} \neq [\mathbf{d}_{m}]_{2}$. For the sake of understandability, we represent the matrix as $\mathbf{D}_{o}=[[\mathbf{d}_{1}]_{1}, [\mathbf{d}_{2}]_{1}; [\mathbf{d}_{1}]_{2}, [\mathbf{d}_{2}]_{2}]$. 
In \cite[Theorem 2]{9188014}, $\mathbf{D}_{2}=[2,1;1,2]$ is proved to be the optimal decoding order from the perspective of providing maximum secrecy rate to each user. Therefore, we consider $\mathbf{D}_{2}$ for our study further. 

For a decoding order, the corresponding achievable data rate at $U_{m}$ can be given by Shannon's capacity formula as
\begin{equation} \label{info_rate}
R_{nm} = \log_{2}(1+\gamma_{nm}),
\end{equation}
where $\gamma_{nm}$ denotes the achievable SINR at $U_{m}$, when $U_{m}$ decodes data of $U_{n}$, where $m, n \in \mathcal{N}$. To analyze the secrecy performance from the perspective of PLS, the achievable secrecy rate $R_{sn}$ at $U_{n}$ can be  expressed as \cite{wyner1975wire}
\begin{equation}\label{secrecy_rate}
R_{sn} = R_{nn} - R_{nm},  \quad n \neq m,
\end{equation}
The condition  $R_{nn}> R_{nm}$ must be satisfied to achieve positive secrecy rate $R_{sn}$ at $U_{n}$.

\section{ Residual Interference Model}\label{section3}

\subsection{Conventional RI Model}
Two models have been proposed in the literature to calculate RI due to imperfect SIC, which are summarized as follows.
\subsubsection{Constant Model}
According to the constant model \cite{9188014}, \cite{constant_sic1}, \cite{constant_sic2}, a certain value of RI is assumed at receivers. Thus, for decoding order $\mathbf{D}_{2}=[2,1;1,2]$, the SINR, $\gamma_{21}$, when $U_2$ is decoded by $U_1$ at the first stage can be expressed as \cite{9188014}
\begin{align}
\gamma_{21} &= \frac{(1-\alpha) |h_{1}|^2}{\alpha |h_{1}|^2+\frac{1}{\rho_{t}}}, \label{sinr21} 
\end{align}
where $\rho_{t}\stackrel{\Delta}{=}\frac{P_{t}}{\sigma^{2}}$ indicates the BS transmit signal-to-noise ratio (SNR). At the second stage of decoding, $U_1$ will receive interference from the previous decoded user $U_2$. Thus, the  SINR $\gamma_{11[C]}$, when $U_1$ decodes data of itself can be given as
\begin{align}
\gamma_{11[C]} &= \frac{\alpha |h_{1}|^2}{\underbrace{\Gamma_{21}}_{\text{RI}}+\frac{1}{\rho_{t}}},  
\end{align}
where $\Gamma_{21}$ is RI from imperfectly decoded signal of $U_2$ treated as an additional noise component.  $[C]$ denotes constant model. Similarly, the SINRs $\gamma_{12}$ and $\gamma_{22[C]} $, respectively, are  given as 
\begin{align}
\gamma_{12} &= \frac{\alpha |h_{2}|^2}{(1-\alpha) |h_{2}|^2+\frac{1}{\rho_{t}}}, \label{sinr12} \\   
\gamma_{22[C]} &= \frac{(1-\alpha) |h_{2}|^2}{\underbrace{\Gamma_{12}}_{\text{RI}}+\frac{1}{\rho_{t}}}, 
\end{align}
where $\Gamma_{12}$ denotes the RI from imperfectly decoded user $U_1$ at the first stage. $\Gamma_{21} = \Gamma_{12} = 0$ represents perfect SIC case.

\subsubsection{Fixed Model}
In the fixed model, the amount of RI is expressed as a fixed fraction of the power of the interfering signal  \cite{9324786}-\cite{8125754}. Since the RI from the imperfectly decoded users is introduced at the second stage of decoding, the SINRs at the first stage, i.e., $\gamma_{21}$ and $\gamma_{12}$, will remain unchanged and will be the same as in \eqref{sinr21} and \eqref{sinr12}, respectively. At the second stage,  $\gamma_{11[F]}$ and  $\gamma_{22[F]}$, respectively, can be expressed as
\begin{align}
\gamma_{11[F]} &= \frac{\alpha |h_{1}|^2}{\underbrace{\beta (1-\alpha) |h_{1}|^2}_{\text{RI}}+\frac{1}{\rho_{t}}},  
\gamma_{22[F]} &= \frac{(1-\alpha) |h_{2}|^2}{\underbrace{\beta \alpha |h_{2}|^2}_{\text{RI}}+\frac{1}{\rho_{t}}}.
\end{align} 
Note that $\beta\in [0,1]$ denotes the fractional residual error coefficient, where $\beta=0$ and $\beta=1$, respectively, represents perfect SIC and no SIC at all. $[F]$ stands for fixed model.

As noted, the constant model, where a certain value of RI is assumed, is over-simplified and might not be reasonable for practical scenarios. For the case of fixed model, the choice of $\beta$ strongly affects the RI. In practice, the actual impacts of imperfect SIC cannot be fully revealed if a fixed value of $\beta$ is adopted. In particular, at the receivers, SIC performance depends significantly on the SINR achieved in the previous stage for decoding the interfering signal's data \cite{8171091}, \cite{8309422}. Therefore, to design a practical system, we next propose a new RI model based on the achievable SINR at decoded users.
 
\subsection{Proposed RI Model}\label{proposed_scheme}
At receivers, whether a signal can be decoded perfectly depends on whether the achievable data rate $R_{nm}$ at $U_{n}$ is larger than the threshold rate $R_{\mathrm{th}}$, where $m, n \in \mathcal{N}$. In terms of SINR, the condition $\gamma_{nm} \geq \gamma_{\mathrm{th}}$ must be satisfied to ascertain that the user can be perfectly decoded. Here, $\gamma_{\mathrm{th}}>0$ represents the SINR threshold for the $n$-th user, where $\gamma_{\mathrm{th}}=2^{R_{\mathrm{th}}}-1$ \cite{9151196}-\cite{7117391}, \cite{9270599}.  As long as this condition holds, the signal is correctly decoded by the receiver. If the rate is lower than the threshold, it means less information rate is achieved to decode the signal successfully. Motivated by this observation, we propose a new generalized RI model.

In the proposed model, firstly, the SINR achieved while cancelling the data of the previous interfering user must be compared with the threshold SINR $\gamma_{\mathrm{th}}$. If the achievable SINR of the user being cancelled is equal to or above the threshold SINR, perfect SIC is considered, which results in zero RI. Otherwise, imperfect SIC is considered, where we quantify the RI by the gap between the threshold SINR required for perfect SIC and the actual achievable SINR at the decoded user in the previous stage. This difference indicates the residual SINR, which was required to decode the signal perfectly.

We consider the two-user untrusted NOMA system with secure decoding order $\mathbf{D}_{2}$, and write the SINRs using the proposed RI model. The achievable SINR $\gamma_{21}$ in the first stage will be the same as in \eqref{sinr21}. In the second stage, the SINR of $U_1$, which experiences the RI from incorrectly decoded user $U_2$ at the first stage can be expressed as
\begin{align}\label{sinrP1}
\gamma_{11[P]}&= 
\begin{cases} 
\gamma^{i}_{11[P]} = \frac{\alpha |h_{1}|^2}{\underbrace{[\gamma_{\mathrm{th}}-\gamma_{21}]\zeta}_{\text{RI}}+\frac{1}{\rho_{t}}}, \quad & \gamma_{21}<\gamma_{\mathrm{th}} \\
\gamma^{p}_{11[P]} = \alpha |h_{1}|^{2} \rho_{t}, & \gamma_{21}\geq\gamma_{\mathrm{th}},
\end{cases} 
\end{align}
where $\gamma^{i}_{11[P]}$ and $\gamma^{p}_{11[P]}$, respectively, denote SINR $\gamma_{11}$ for imperfect and perfect SIC case. Here, to normalize RI, we have used a sensitivity parameter $\zeta$, which not only converts the expression into the desired form, but also acts as a controller that decides the severity of the SIC error. The value of $\zeta$ depends on the underlying applications. The higher value of $\zeta$ is suited for the applications that are critical to SIC so that no penalty is allowed. $\zeta$ can be selected as a lower value for noise-limited applications where the impact of SIC is negligible. $[P]$ indicates the proposed RI model. Similarly, for $U_{2}$, the achievable SINR $\gamma_{12}$ at the first stage will be the same as given in \eqref{sinr12}, and the SINR $\gamma_{11[P]}$ having the RI from imperfectly decoded user $U_1$ can be written as
\begin{align}\label{sinrP2}
\gamma_{22[P]}&= 
\begin{cases} 
\gamma^{i}_{22[P]} = \frac{(1-\alpha) |h_{2}|^2}{\underbrace{[\gamma_{\mathrm{th}}-\gamma_{12}]\zeta}_{\text{RI}}+\frac{1}{\rho_{t}}}, \quad & \gamma_{12}<\gamma_{\mathrm{th}} \\
\gamma^{p}_{22[P]} = (1-\alpha) |h_{2}|^{2} \rho_{t}, & \gamma_{12}\geq\gamma_{\mathrm{th}}
\end{cases} 
\end{align}
where $\gamma^{i}_{22[P]}$ and $\gamma^{p}_{22[P]}$, represent SINR for imperfect and perfect SIC case, respectively, when $U_2$ decodes its own data.




\section{Secrecy Performance Analysis}\label{section5}
\subsection{Exact SOP Analysis}
Considering the RI at receivers with the proposed RI model, now, we derive analytical expressions of SOP for both users. SOP for a user is defined as the probability that the maximum achievable secrecy rate at the user is lesser than a target secrecy rate. Let us denote SOP for $U_{n}$ by $S_{n}$, where $n \in \mathcal{N}$. 
\subsubsection{Strong User}
Using \eqref{sinrP1}, we consider the probability of occurrence of imperfect and perfect SIC, respectively, as $\mathbb{P}\{\gamma_{21}<\gamma_{\mathrm{th}}\}$ and $\mathbb{P}\{\gamma_{21}\geq\gamma_{\mathrm{th}}\}$, and thus, the SOP for $U_{1}$, i.e., $S_{1}$, can be mathematically expressed as 
\begin{align}\label{SOP_1}
S_{1} \!&=\! \mathbb{P}\{\gamma_{21}<\gamma_{\mathrm{th}}\} S^{i}_{1} + \mathbb{P}\{\gamma_{21}\geq\gamma_{\mathrm{th}}\} S^{p}_{1} ,\nonumber \\
&= \mathbb{P}\Big\{ |h_{1}|^{2} Z_{1} < \gamma_{\mathrm{th}} \Big\}  S^{i}_{1} + \mathbb{P}\Big\{ |h_{1}|^{2} Z_{1} \geq  \gamma_{\mathrm{th}}  \Big\} S^{p}_{1}, \nonumber \\
&=\!\!
\begin{cases}
\mathbb{P}\Big\{ |h_{1}|^{2} \! < \! \frac{\gamma_{\mathrm{th}}}{Z_{1}} \Big\} S^{i}_{1}  +  \mathbb{P}\Big\{ |h_{1}|^{2} \! \geq \! \frac{\gamma_{\mathrm{th}}}{Z_{1}} \Big\} S^{p}_{1}, \! &  \!  Z_{1} \geq 0 \\
\mathbb{P}\Big\{ |h_{1}|^{2} \! > \! \frac{\gamma_{\mathrm{th}}}{Z_{1}} \Big\} S^{i}_{1}  +  \mathbb{P}\Big\{ |h_{1}|^{2} \! \leq \! \frac{\gamma_{\mathrm{th}}}{Z_{1}} \Big\} S^{p}_{1}, \! & \!\!\! \text{otherwise}
\end{cases} \nonumber \\
\!&=\!\!
\begin{cases}\!
\Big(1\!-\!\exp\!\Big\{\frac{-\gamma_{\mathrm{th}}}{Z_{1}\lambda_{1}}\Big\}\!\Big) S^{i}_{1} \! + \!  \exp\Big\{\frac{-\gamma_{\mathrm{th}}}{Z_{1}\lambda_{1}}\Big\} S^{p}_{1}, \!  & \!\!\! \alpha \! \leq \! \frac{1}{1+\gamma_{\mathrm{th}}} \\
1 \times S^{i}_{1} + 0 \times S^{p}_{1}, &\!\!\!\!\! \text{otherwise}
\end{cases}
\end{align}
where $\mathbb{P}\{.\}$ denotes the probability measure, $S^{i}_{1}$ and $S^{p}_{1}$, respectively, represent the SOP expressions at $U_{1}$ for imperfect and perfect SIC case, and $Z_{1}=((1-\alpha)-\alpha\gamma_{\mathrm{th}})\rho_{t}$. For imperfect SIC case,  considering $R^{i}_{s1}$ and $R_{s1}^{\mathrm{th}}$, respectively, as the achievable secrecy rate and target secrecy rate for $U_{1}$, and $\Pi_{1} \stackrel{\Delta}{=} 2^{R_{s1}^{\mathrm{th}}}$, $S^{i}_{1} $, using \eqref{secrecy_rate}, \eqref{sinr12}, and \eqref{sinrP1}, can be obtained as
\begin{align}\label{SOP_1i}
S^{i}_{1} &= \mathbb{P}\{R^{i}_{s1} <  R^{\mathrm{th}}_{s1}\} =\mathbb{P}\Bigg\{\frac{1+\gamma^{i}_{11[P]}}{1+\gamma_{12}} < \Pi_{1}\Bigg\}, \nonumber \\
&= \mathbb{P}\Big\{A_{1}(|h_{1}|^{2})^{2} + B_{1} |h_{1}|^{2} + C_{1} < 0\Big\}, \nonumber \\
&= \mathbb{P}\{V_{1}<|h_{1}|^{2}<W_{1}\}, \nonumber \\
&= \int_{0}^{\infty} ( F_{|h_{1}|^{2}}(W_{1}) - F_{|h_{1}|^{2}}(V_{1}) ) f_{| h_{2} |^{2}}(y_{1}) dy_{1},\nonumber \\
&= \frac{1}{\lambda_{2}} \int_{0}^{\infty} \bigg(1 - \exp\bigg\{\frac{-W_{1}}{\lambda_{1}}\bigg\}\bigg)  \exp\bigg\{\frac{-y_{1}}{\lambda_{2}}\bigg\}dy_{1}.
\end{align}
In \eqref{SOP_1i}, $A_{1}$, $B_{1}$, and $C_{1}$ are given as
\begin{align}\label{A1B1C1}
A_{1}&=\alpha^{2}\rho_{t}^{2}K_{1}, \nonumber \\
B_{1}&=(\alpha  + (\Pi_{1}-1) (1-\alpha) \zeta \rho_{t}  - (\Pi_{1}-1) \gamma_{\mathrm{th}} \alpha \zeta \rho_{t} \nonumber \\
 & - (\Pi_{1}-1)\alpha)K_{1}\rho_{t} + ((1-\alpha) \zeta \rho_{t}  -   \alpha \gamma_{\mathrm{th}} \zeta \rho_{t} \nonumber \\
 &  - \alpha)\Pi_{1}|h_{2}|^{2} \alpha \rho_{t}^{2}, \nonumber \\
C_{1}&=(- (\Pi_{1}-1)  \gamma_{\mathrm{th}}\zeta \rho_{t}  - (\Pi_{1}-1) )K_{1} \nonumber \\
& - \Pi_{1}|h_{2}|^{2} \alpha \gamma_{\mathrm{th}} \zeta \rho_{t}^{2}   - \Pi_{1}|h_{2}|^{2}\alpha\rho_{t},
\end{align}
where $K_{1}=(1-\alpha)|h_{2}|^{2}\rho_{t}+1$. Then, $V_{1} = \frac{-B_{1} - \sqrt{B_{1}^{2} - 4A_{1}C_{1}}}{2A_{1}}$ and $W_{1} = \frac{-B_{1} + \sqrt{B_{1}^{2} - 4A_{1}C_{1}}}{2A_{1}}$. $F_{|h_{1}|^{2}}(X)$ and $f_{\mid h_{2} \mid ^{2}}(X)$, respectively, denote the cumulative distribution function  and the probability density function of channel power gain $|h_{1}|^{2}$ and $|h_{2}|^{2}$. Here $A_{1}>0$, which indicates $W_{1}>V_{1}$. Since  $C_{1}<0$, $V_{1}<0$ implies $F_{|h_{1}|^{2}}(V_{1})=0$ because the exponential distribution is supported in the interval $[0,\infty)$. 

Likewise, the expression of SOP for perfect SIC case, i.e., $S^{p}_{1}$ can be obtained, which has already been derived in \cite{globecom} .
\begin{Remark}\label{remark2}
From \eqref{SOP_1}, we observe that SOP for $U_{1}$ depends on the probability of occurrence of imperfect and perfect SIC. Also, from \eqref{sinrP1}, we note that in perfect SIC case, the value of RI is zero, while in imperfect SIC case, RI depends significantly on the gap between $\gamma_{\mathrm{th}}$ and $\gamma_{21}$. Thus, for given system parameters, where  $\gamma_{21}<\gamma_{\mathrm{th}}$, if we further increase the  $\gamma_{\mathrm{th}}$, the RI increases due to which $\gamma^{i}_{11[P]}$ decreases. As a result, the secrecy rate for $U_{1}$ decreases, and hence SOP increases. 
\end{Remark}
\subsubsection{Weak User}
Going on similar lines, the SOP expression for $U_{2}$, i.e., $S_{2}$, by using \eqref{sinrP2}, can be given as 
\begin{align}\label{SOP_2}
S_{2} &= \mathbb{P}\{\gamma_{12}<\gamma_{\mathrm{th}}\} S^{i}_{2} + \mathbb{P}\{\gamma_{12}\geq\gamma_{\mathrm{th}}\} S^{p}_{2} ,\nonumber \\
\!&=\!\!
\begin{cases}\!
\Big(1\!-\!\exp\!\Big\{\frac{-\gamma_{\mathrm{th}}}{Z_{2}\lambda_{2}}\Big\}\!\Big) S^{i}_{2} \! + \!  \exp\Big\{\frac{-\gamma_{\mathrm{th}}}{Z_{2}\lambda_{2}}\Big\} S^{p}_{2}, \!  & \!\!\! \alpha \! \geq \! \frac{\gamma_{\mathrm{th}}}{1+\gamma_{\mathrm{th}}} \\
1 \times S^{i}_{2} + 0 \times S^{p}_{2}, &\!\!\!\!\! \text{otherwise.}
\end{cases}
\end{align}
Here $Z_{2}=(\alpha-(1-\alpha)\gamma_{\mathrm{th}})\rho_{t}$. $S^{i}_{2}$ and $S^{p}_{2}$, respectively, indicate the SOP expressions for imperfect and perfect SIC case. For the case of imperfect SIC, let us consider $R^{i}_{s2}$ as the achievable secrecy rate and $R_{s2}^{\mathrm{th}}$ as the target secrecy rate for $U_{2}$. Using \eqref{secrecy_rate}, \eqref{sinr21}, and \eqref{sinrP2}, $S^{i}_{2} $ can be expressed as
\begin{align}\label{SOP_2i}
S^{i}_{2} &= \mathbb{P}\{R^{i}_{s2} <  R_{s2}^{\mathrm{th}}\} = \mathbb{P}\Bigg\{\frac{1+\gamma^{i}_{22[P]}}{1+\gamma_{21}} < \Pi_{2}\Bigg\}, \nonumber \\
&= \frac{1}{\lambda_{1}} \int_{0}^{\infty} \bigg(1 - \exp\bigg\{\frac{-W_{2}}{\lambda_{2}}\bigg\}\bigg) \exp\bigg\{\frac{-y_{2}}{\lambda_{1}}\bigg\}dy_{2}.
\end{align}
In \eqref{SOP_2i}, $\Pi_{2} \stackrel{\Delta}{=} 2^{R_{s2}^{\mathrm{th}}}$, $W_{2} = \frac{-B_{2} + \sqrt{B_{2}^{2} - 4A_{2}C_{2}}}{2A_{2}}$, where $A_{2}$, $B_{2}$, and $C_{2}$ are given as
\begin{align}\label{A2B2C2}
A_{2}&=(1-\alpha)^{2}\rho_{t}^{2} K_{2}, \nonumber \\
B_{2}&=((1-\alpha) + (\Pi_{2}-1) \alpha \zeta \rho_{t} 
 - (\Pi_{2}-1) \gamma_{\mathrm{th}} (1-\alpha) \zeta \rho_{t}  \nonumber \\ &- (\Pi_{2}-1) (1-\alpha))K_{2}\rho_{t}  + (\alpha \zeta \rho_{t}  -  (1-\alpha)\gamma_{\mathrm{th}} \zeta \rho_{t}  \nonumber \\
& -  (1-\alpha))\Pi_{2}|h_{1}|^{2} (1-\alpha) \rho_{t}^{2}, \nonumber \\
C_{2} &=(- (\Pi_{2}-1)  \gamma_{\mathrm{th}}\zeta \rho_{t}  - (\Pi_{2}-1)) K_{2} \nonumber \\
& - \Pi_{2}|h_{1}|^{2} (1-\alpha) \gamma_{\mathrm{th}} \zeta \rho_{t}^{2}  - \Pi_{2}|h_{1}|^{2}(1-\alpha)\rho_{t}, 
\end{align}
with $K_{2}=\alpha|h_{1}|^{2}\rho_{t}+1$. 
Similarly, we can obtain the SOP expression $S^{p}_{2}$ for perfect SIC case,  which is provided in  \cite{globecom}.
\begin{Remark}\label{remark1}
From \eqref{SOP_1i} and \eqref{SOP_2i} as well as from the SOP expressions given in \cite{globecom} for perfect SIC case, we can observe that SOPs for $U_{1}$ and $U_{2}$ depend on $R_{s1}^{\mathrm{th}}$ and $R_{s2}^{\mathrm{th}}$, respectively. With the reason that an outage occurs when the achievable secrecy rate at a user falls underneath a target rate, the SOP increases on increasing the target secrecy rate at a user.
 \end{Remark}
 
\subsection{Asymptotic Approximations}
The analytical expressions of SOPs are challenging to solve. Therefore, to provide  deeper insights on the obtained results, we present their high SNR closed-form approximations. Let us represent asymptotic SOP for $U_{n}$ by $\widehat S_{n}$, where $n \in \mathcal{N}$. 

\subsubsection{Strong User}
The closed-form high SNR approximation of $S_{1}$ given in \eqref{SOP_1}, i.e., $\widehat S_{1}$, can be written as 
\begin{align}\label{SOP_asy_1}
\widehat S_{1} &= \mathbb{P}\{\gamma_{21}<\gamma_{\mathrm{th}}\} \widehat S^{i}_{1} + \mathbb{P}\{\gamma_{21}\geq\gamma_{\mathrm{th}}\} \widehat S^{p}_{1} ,\nonumber \\
&=
\begin{cases}\!
\Big(1\!-\!\exp\!\Big\{\frac{-\gamma_{\mathrm{th}}}{Z_{1}\lambda_{1}}\Big\}\!\Big) \widehat S^{i}_{1} +   \exp\Big\{\frac{-\gamma_{\mathrm{th}}}{Z_{1}\lambda_{1}}\Big\} \widehat S^{p}_{1}, \!  & \!\!\! \alpha \! \leq \! \frac{1}{1+\gamma_{\mathrm{th}}} \\
1 \times \widehat S^{i}_{1} + 0 \times \widehat S^{p}_{1}, &\!\!\!\!\! \text{otherwise.}
\end{cases}
\end{align}
Here, $\widehat S^{i}_{1}$ and $\widehat S^{p}_{1}$, are asymptotic SOP expressions at $U_{1}$ for imperfect and perfect SIC case, respectively, which can be obtained for $\rho_{t}\gg1$. For imperfect SIC case, by using $K_1 = ((1-\alpha)|h_{2}|^{2}\rho_{t}+1)$ $\approx$ $(1-\alpha)|h_{2}|^{2}\rho_{t}$ in \eqref{A1B1C1}, the high SNR approximation of $S^{i}_{1}$ given in \eqref{SOP_1i}, i.e., $\widehat S^{i}_{1}$ can be obtained as
\begin{align} \label{sop1_asy_i}
\widehat S^{i}_{1} = 1 - \exp\bigg\{\frac{-\widehat W_{1}} {\lambda_{1}}\bigg\}.
\end{align}
In \eqref{sop1_asy_i}, $\widehat W_{1}=\frac{-\widehat B_{1} + \sqrt{\widehat B_{1}^{2} - 4\widehat A_{1} \widehat C_{1}}}{2 \widehat A_{1}}$, where 
\begin{align}\label{A1B1C1_asy}
\widehat A_{1}&=\alpha^{2}\rho_{t}^{3} (1-\alpha), \nonumber \\
\widehat B_{1}&=(\alpha  + (\Pi_{1}-1) (1-\alpha) \zeta \rho_{t}  - (\Pi_{1}-1) \gamma_{\mathrm{th}} \alpha \zeta \rho_{t} \nonumber \\
 & - (\Pi_{1}-1)\alpha)(1-\alpha)\rho_{t}^{2} + ((1-\alpha) \zeta \rho_{t}  -   \alpha \gamma_{\mathrm{th}} \zeta \rho_{t} \nonumber \\
 &  - \alpha)\Pi_{1} \alpha \rho_{t}^{2}, \nonumber \\
C_{1}&=(- (\Pi_{1}-1)  \gamma_{\mathrm{th}}\zeta \rho_{t}  - (\Pi_{1}-1) )(1-\alpha)\rho_{t} \nonumber \\
& - \Pi_{1} \alpha \gamma_{\mathrm{th}} \zeta \rho_{t}^{2}   - \Pi_{1}\alpha\rho_{t}.
\end{align}
In a similar manner, $\widehat S^{p}_{1}$ has been derived in \cite{globecom}.

\subsubsection{Weak User}
The closed-form high SNR approximation of $S_{2}$ given in \eqref{SOP_2}, i.e., $\widehat S_{2}$, can be expressed as
\begin{align}\label{SOP_asy_2}
\widehat S_{2} &= \mathbb{P}\{\gamma_{12}<\gamma_{\mathrm{th}}\} \widehat S^{i}_{2} + \mathbb{P}\{\gamma_{12}\geq\gamma_{\mathrm{th}}\} \widehat S^{p}_{2} ,\nonumber \\
\!&=\!\!
\begin{cases}\!
\Big(1\!-\!\exp\!\Big\{\frac{-\gamma_{\mathrm{th}}}{Z_{2}\lambda_{2}}\Big\}\!\Big) \widehat S^{i}_{2} \! + \!  \exp\Big\{\frac{-\gamma_{\mathrm{th}}}{Z_{2}\lambda_{2}}\Big\} \widehat S^{p}_{2}, \!  & \!\!\! \alpha \! \geq \! \frac{\gamma_{\mathrm{th}}}{1+\gamma_{\mathrm{th}}} \\
1 \times \widehat S^{i}_{2} + 0 \times \widehat S^{p}_{2}, &\!\!\!\!\! \text{otherwise.}
\end{cases}
\end{align}
In \eqref{SOP_asy_2}, $\widehat S^{i}_{2}$ and $\widehat S^{p}_{2}$,  respectively, are high SNR SOP approximations at $U_{2}$ for imperfect and perfect SIC case. For imperfect SIC case, by setting $K_2 = (\alpha|h_{1}|^{2}\rho_{t}+1)$ $\approx$ $\alpha|h_{1}|^{2}\rho_{t}$ in \eqref{A2B2C2}, the obtained $\widehat S^{i}_{2}$  can be written as
\begin{align} \label{sop2_asy_i}
\widehat S^{i}_{2} = 1 - \exp\bigg\{\frac{-\widehat W_{2}} {\lambda_{2}}\bigg\}.
\end{align}
In \eqref{sop2_asy_i}, $\widehat W_{2}=\frac{-\widehat B_{2} + \sqrt{\widehat B_{2}^{2} - 4\widehat A_{2} \widehat C_{2}}}{2 \widehat A_{2}}$, where 
\begin{align}\label{A2B2C2_asy}
\widehat A_{2}&=(1-\alpha)^{2}\rho_{t}^{3} \alpha, \nonumber \\
\widehat B_{2}&=((1-\alpha) + (\Pi_{2}-1) \alpha \zeta \rho_{t} 
 - (\Pi_{2}-1) \gamma_{\mathrm{th}} (1-\alpha) \zeta \rho_{t}  \nonumber \\ &- (\Pi_{2}-1) (1-\alpha))\alpha\rho_{t}^{2}  + (\alpha \zeta \rho_{t}  -  (1-\alpha)\gamma_{\mathrm{th}} \zeta \rho_{t}  \nonumber \\
& -  (1-\alpha))\Pi_{2} (1-\alpha) \rho_{t}^{2}, \nonumber \\
C_{2} &=(- (\Pi_{2}-1)  \gamma_{\mathrm{th}}\zeta \rho_{t}  - (\Pi_{2}-1)) \alpha\rho_{t} \nonumber \\
& - \Pi_{2} (1-\alpha) \gamma_{\mathrm{th}} \zeta \rho_{t}^{2}  - \Pi_{2}(1-\alpha)\rho_{t}.
\end{align}
Likewise, the asymptotic  SOP $\widehat S^{p}_{2}$ has been derived in \cite{globecom}.

\section{Numerical Investigation}\label{section7}

\begin{figure}[!t]
\centering
\includegraphics[scale=.37]{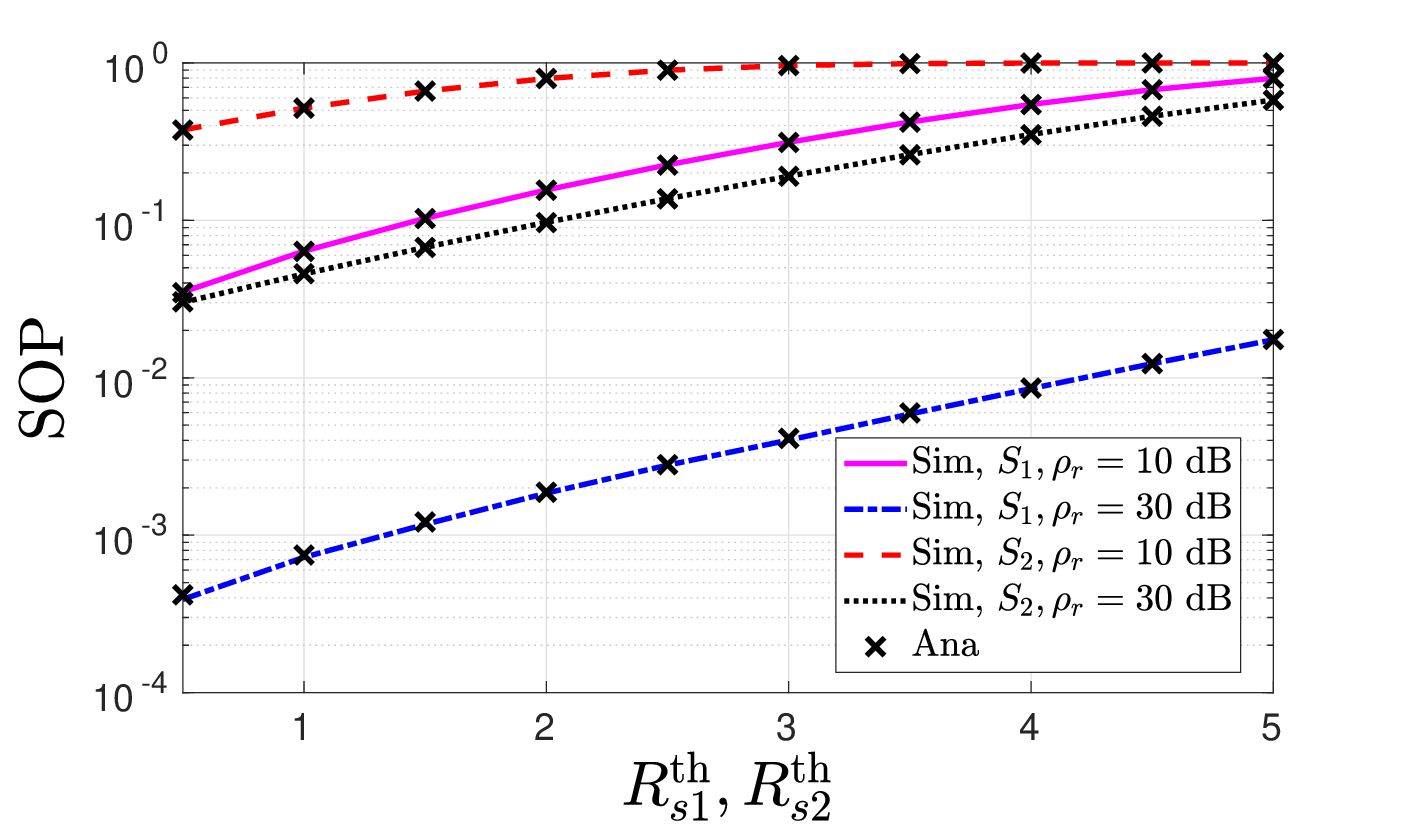}
\caption{Validation of SOPs for both users $U_{1}$ and $U_{2}$ with $\alpha=0.33$}
\label{validation_sim_ana}
\end{figure}
\begin{figure}[!t]
\centering
\includegraphics[scale=.37]{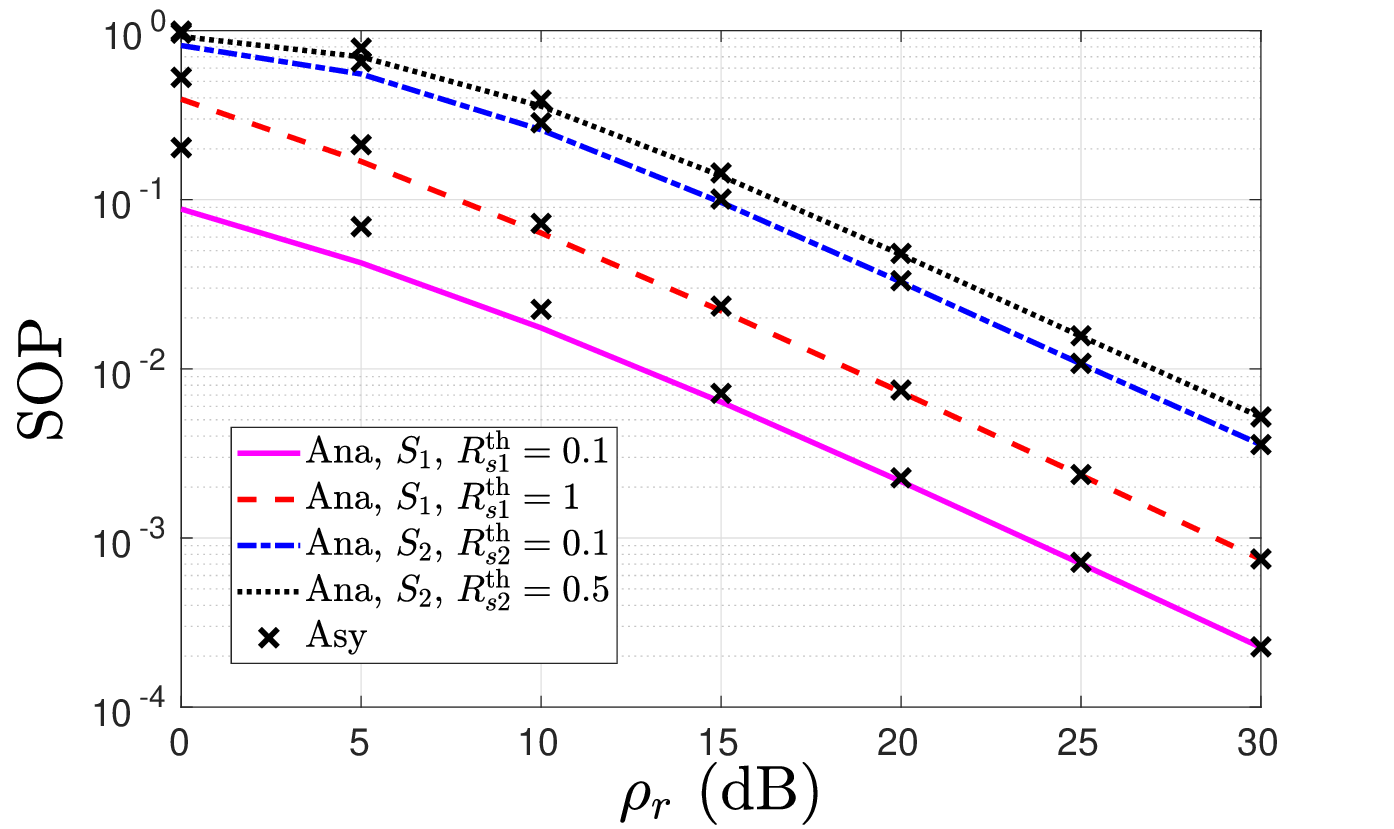}
\caption{Validation of analytical and asymptotic approximation of SOPs for both users $U_{1}$ and $U_{2}$ with $\alpha=0.33$.}
\label{validation_ana_asy}
\end{figure}

We present numerical results to validate the derived analysis. We have used $d_{1}=50$ and $d_{2}=100$ meters. Noise power is set to $-90$ dBm with the noise signal following Gaussian distribution at all users. Small scale fading is considered to follow the Rayleigh distribution with mean value $1$. Simulations are averaged over $10^6$ randomly generated channel realizations using Rayleigh distribution at both users. Besides, $L_{c}=1$, $e=3$, $\zeta=10^{-10}$, $\gamma_{\mathrm{th}}=1$,  $\rho_{t}=70$ dB, $\alpha=0.5$ and $R_{s1}^{\mathrm{th}}=R_{s2}^{\mathrm{th}}=0.1$ are taken. $\rho_{r}$ denotes the received SNR in decibels (dB) at $U_{2}$. Simulation, analytical, and asymptotic results are, respectively, marked as `Sim', `Ana', and `Asy'.

\subsection{Validation of Analysis}
In Fig. \ref{validation_sim_ana}, for different values of $\rho_{r}$, the validation of SOPs, $S_{1}$ with $R_{s1}^{\mathrm{th}}$ and $S_{2}$ with $R_{s2}^{\mathrm{th}}$  are shown. The perfect match between simulated and analytical results shows the correctness of the analysis. It can be observed from the results that $S_{1}$ and $S_{2}$ increase with $R_{s1}^{\mathrm{th}}$ and $R_{s2}^{\mathrm{th}}$, respectively.  This is because with an increase in target rate, outage increases. We also observed that with an increase in $\rho_{r}$, both $S_{1}$ and $S_{2}$ decrease. The reason is secrecy rates achieved at a user increases with an increase in SNR, and thus, for a given target secrecy rate, SOP decreases. Besides, from Fig. \ref{validation_ana_asy}, it can be clearly seen that analytical results match with asymptotic results at high SNR, which confirms the accuracy of asymptotic expressions.

\subsection{Impact of Relative Distance between Users}
\begin{figure}[!t]
\centering
\includegraphics[scale=.36]{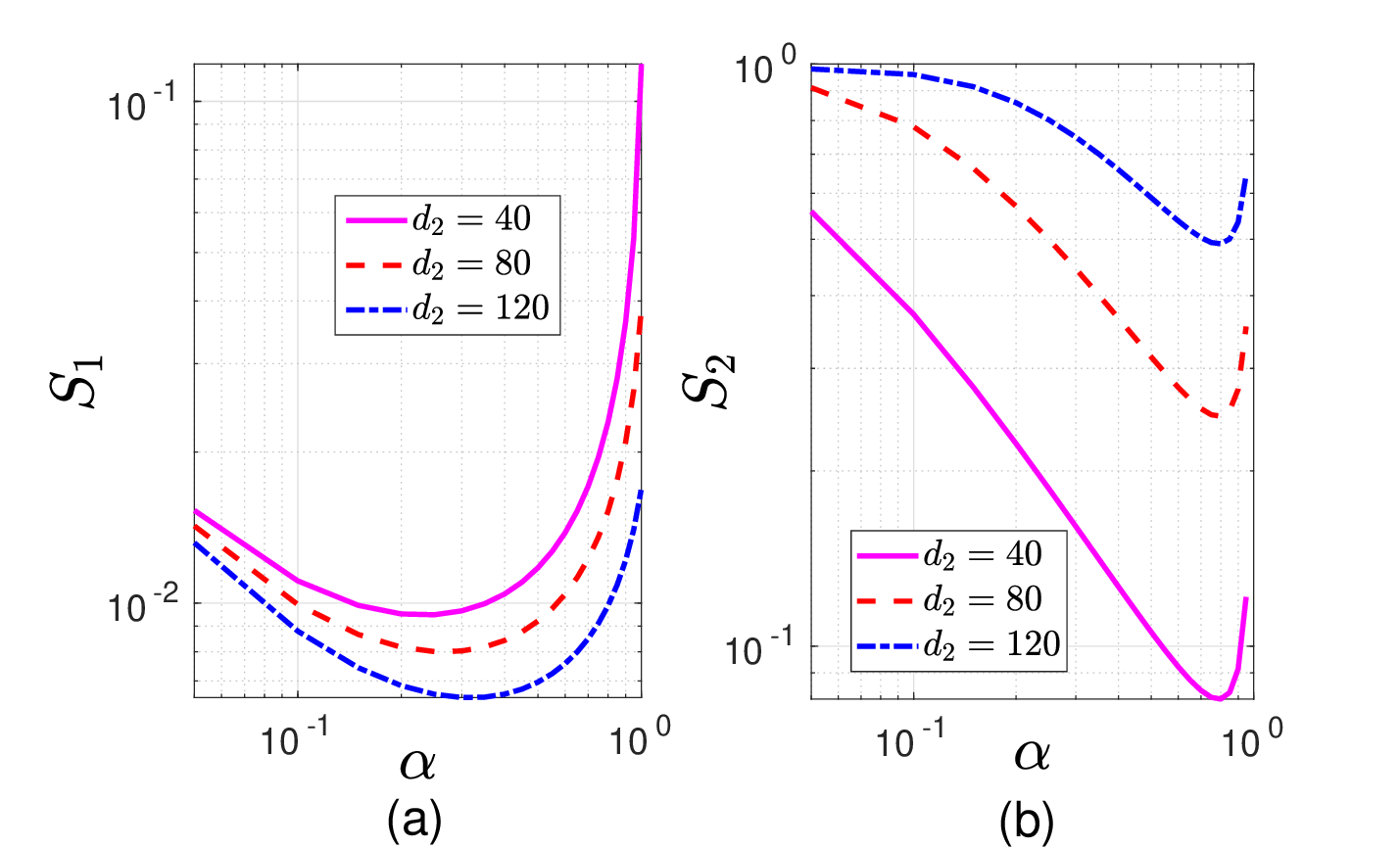}
\caption{Insights on optimal power allocation $\alpha$ that minimizes $S_1$ in (a) and $S_2$ in (b), for different values of $d_{2}$, with $d_{1}=40$ meter.}
\label{validation_optimal_PA}
\end{figure}

Fig. \ref{validation_optimal_PA}(a) is plotted to observe the impact of varying the distance $d_{2}$ on achievable SOPs, where $d_{1}=40$ meters. We notice that $S_{1}$ decreases with an increase in $d_{2}$. This reason is an increase in $d_{2}$ causes a drop in achievable data rate at $U_{2}$, which provides a better secrecy rate at $U_{1}$. On the contrary, as shown in Fig. \ref{validation_optimal_PA}(b), due to a decrease in data rate at $U_{2}$, secrecy rate at $U_{2}$ decreases and thus, SOP increases.  In conclusion, it can be said that increasing the distance from BS to $U_{2}$ has a contradictory effect on $S_{1}$ and $S_{2}$. Besides, from Fig. \ref{validation_optimal_PA}(a) and Fig.\ref{validation_optimal_PA}(b), we notice that there exist an optimal PA that minimizes the SOP performance for both users $U_{1}$ and $U_{2}$. 

\subsection{Insights on the effect of Sensitivity Parameter}
\begin{figure}[!t]
\centering
\includegraphics[scale=.37]{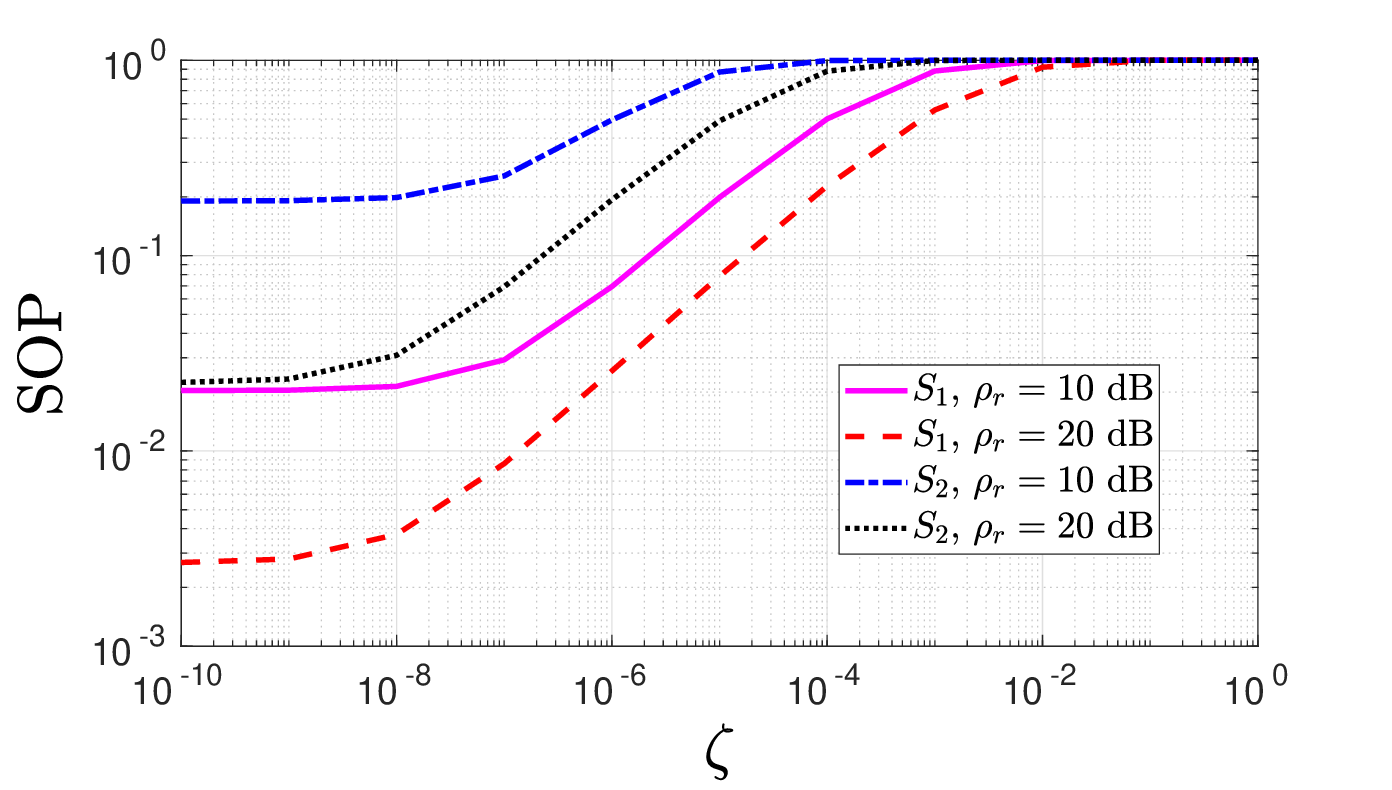}
\caption{Variation of SOPs for both users, $U_{1}$ and $U_{2}$, with sensitivity parameter $\zeta$ for different values of $\rho_{r}$.}
\label{senstivity_parameter}
\end{figure}

In Fig. \ref{senstivity_parameter} is plotted to study the effect of the sensitivity parameter $\zeta$ on SOPs for both users. We observe that $S_{1}$ and $S_{2}$ increase with the increase in $\zeta$. This is because, with an increase in $\zeta$, less data rate is obtained at a user resulting in a decrease in its secrecy rate and an increase in the SOP. It is also observed that as the value of $\zeta$ increases, the gap between achievable SOPs for two different SNR values minimize. It shows that when less value of $\zeta$ is taken, SOP depends significantly on SNR. Conversely, the high values of $\zeta$ indicate that SOPs is strongly dependent on the interfering signals. Thus, it can be concluded that for noise-limited applications, a lesser value of $\zeta$ should be taken, while for interference-limited applications, higher values of $\zeta$ should be considered.

\begin{figure}[!t]
\centering
 \includegraphics[scale=.37]{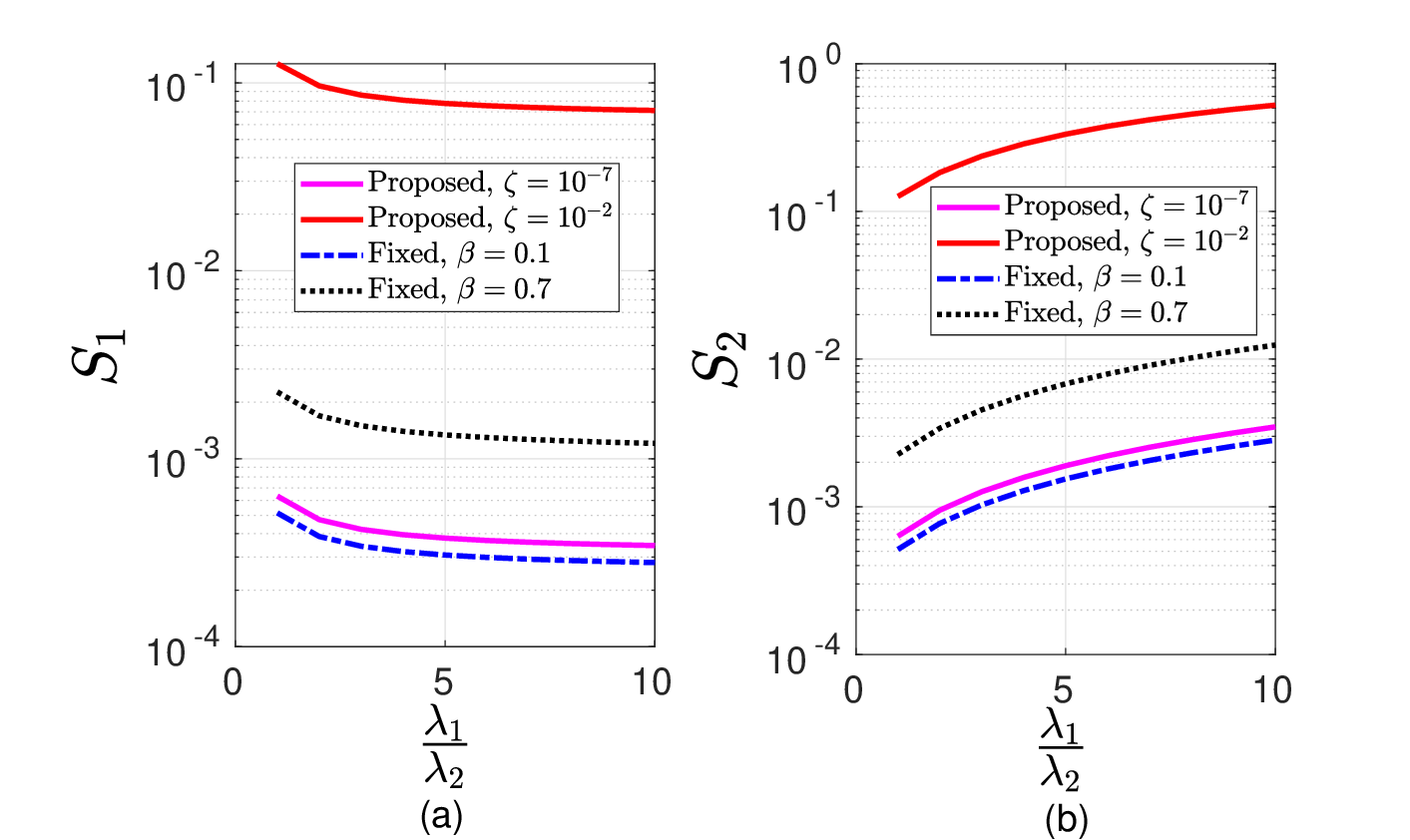}
\caption{Impact on $S_{1}$ in (a) and $S_{2}$ in (b) with relative channel power gain conditions between BS and users, for the proposed and the fixed RI model.}
\label{performance}
\end{figure}

\subsection{Impact of Relative Channel Power Gain Conditions}
Fig. \ref{performance} presents the impact of relative channel power gain conditions between BS and users on derived SOPs for both the proposed and fixed RI model. The analytical expressions of SOPs for the fixed model are provided in \cite{eusipco}. We observe that in both the models, there is a contradictory effect on $S_{1}$ and $S_{2}$ with an increase in $\frac{\lambda_{1}}{\lambda_{2}}$. It means in order to improve the secrecy of a strong user $U_{1}$, $|h_{1}|^{2}\gg|h_{2}|^{2}$. Conversely, better secrecy performance is obtained for a weak user if the channel power gain difference is less. The reason is when $|h_{1}|^{2}\gg|h_{2}|^{2}$, a higher information rate is achieved at the strong user to decode its own data, and the lesser rate is obtained at the weak user to decode strong user's data, resulting in a better secrecy rate at strong user due to which SOP decreases.

\section{Conclusions}\label{section8}
In this work, we have proposed a new generalized RI model to analyze the realistic SIC effect at practical receivers. Observing RI at receivers based on the proposed model in a two-user untrusted NOMA system, we have derived the analytical and asymptotic expressions of SOP for both users. Numerical results have validated the derived mathematical expressions and have provided insights into the impact of various parameters on system performance. Future work includes extending the study of secrecy in a multi-user NOMA scenario.

\bibliographystyle{IEEEtran}
\bibliography{ref}
\end{document}